\documentstyle[12pt]{article}
 2
\begin{document}

\title{ {\normalsize{\hfill SNB/96/1}}\\
\vspace{-.5cm}
{\normalsize{\hfill May,1996}}{\vskip 0.5cm}
\bf FERMIONS ON LATTICE AND CHIRAL INVARIANCE}
\vskip 1in
\author{
{\bf\it H. Banerjee\thanks{Invited lecture at the International
Conference on ``70 years of Quantum Mechanics and Recent Trends
in Theoretical Physics'', ISI, Calcutta Jan.29 - Feb.2, 1996. To
be published in the proceedings (World Scientific, Singapore,
1996).} $^{a}$ {\em and}  Asit K. De $^{b}$}\\
{\small $^a$ S.N.Bose National Centre For Basic Sciences}\\
{\small Block-JD, Sector-III, Salt Lake, Calcutta-700091, India}\\
{\small e-mail: banerjee@bose.ernet.in}\\
{\small $^b$ Saha Institute of Nuclear Physics}\\
{\small AF-Block, Sector I, Salt Lake Calcutta - 700064, India}\\
{\small e-mail : de@tnp.saha.ernet.in}
}
\date{}
\newpage
\maketitle
\begin{abstract}
A model for lattice fermion is proposed which is, (i) free from
doublers, (ii) hermitian, and (iii) chirally invariant. The price paid
is the loss of hypercubic  and reflection symmetries in the lattice
action. Thanks to the $\epsilon$-prescription, correlation functions
are free from the ill effects due to the loss of these symmetries. In
weak coupling approximation, the U(1) vector current of a gauge theory
of lattice fermion in this model is conserved in the continuum limit.
As for the U(1) axial vector current, one obtains the ABJ anomaly if
the continuum limit is implemented before the chiral limit $m = 0$.
The anomaly disappears, as in the Wilson model, if the order of the
two limits is reversed.
\end{abstract}
\newpage
\section{Introduction}

Lattice formulation provides a framework for quantum field theory
where one can, in principle, address questions which are
non-perturbative. Interacting gauge fields are easily
incorporated through the link variables and the lattice spacing `$a$'
provides built-in gauge-invariant regularisation. In practical
applications, however, a serious impediment is the problem of
fermion doubling$^{1}$. The `naive' lattice action
leads$^{1}$ to fifteen `doublers' all with the same mass as the
physical fermion and survive in the spectrum in the continuum
limit.

A `no-go' theorem$^{2}$ states that the unwanted doublers can be
evaded, but only at the price of some basic, desirable
properties$^{3}$ of lattice fermion action. A variety of models
have appeared in literature which get rid of the doublers, either
partially or completely, by abandoning one or more of these basic
properties. The most popular among these, the Wilson model,
abandons chiral symmetry$^{1,4}$. The Wilson term in the lattice
action of this model vanishes formally in the continuum limit, i.e.
is an irrelevant term, and breaks chiral symmetry explicitly. It
does the job of removing all the doublers by giving each of them a
mass of $0\left({1/a}\right)$, i.e., of the order of the cut-off. For a
vector-like gauge theory, e.g., QCD, the Wilson model is 
adequate. In weak coupling perturbation theory ({\it wcpt}) it reproduces
in the continuum limit the correct Ward identities, and, most
importantly, the ABJ anomaly in the U(1) axial current$^{1,5}$. The
explicit breaking of chiral symmetry, however, renders the Wilson
model unsuitable for chiral gauge theories like the Standard
Model. The difficulty with chiral gauge theory is not specific for
the Wilson model, but an outstanding problem of lattice formulation
and quantum field theory in general.

It is clear that in order to be compatible with chiral gauge
theories the lattice action for fermion should have exact chiral
symmetry. Even the irrelevant terms, like the Wilson term, needed
to remove the doublers from the spectrum, must leave chiral
symmetry unscathed. In view of the no-go theorem, it is also
clear that the chirally symmetric irrelevant term must pay some
price. What, however, is not clear is whether the model will still
be physically acceptable. A decisive criterion for this would be
the Ward identities in {\it wcpt}. We report here the results of a search
for a chirally symmetric lattice fermion action which satisfies
this criterion.

In sect.2 we introduce various options for finite differences on
lattice, all of which are candidates for derivative in continuum.
The specific choice of finite difference for our lattice action
(sect.3) is motivated by the three principles, (i) absence of
doublers, (ii) strict chiral symmetry, and (iii) hermiticity.
The price paid$^{3}$ is the loss of
hypercubic and reflection symmetries$^{6,7}$. We, however, give
a prescription to remedy the lack of these symmetries in
correlation functions. In sect.4 we derive the Ward identities
of the U(1) curents in {\it wcpt} and show that the vector current is
conserved while the axial current has the ABJ anomaly in the
continuum limit. Important steps in the derivation of these
identities are given in the Appendix. We conclude with some remarks
in sect.5.

\section{Finite Differences and Fermion Doublers}

There are several options for finite differences on lattice, all of
which coincide with derivative in the continuum limit :
\\ \\
(a) forward difference
$$\delta^{f}_{\mu} \equiv \frac{e^{-ip_{\mu}a} - 1}{a} \equiv -
iP_{\mu} \eqno(2.1)$$
(b) backward difference
$$\delta^{b}_{\mu} \equiv \frac{1 - e^{ip_{\mu}a}}{a}  \equiv -
iP_{\mu}^{+} \eqno(2.2)$$
(c) symmetric difference
$$\delta^{s}_{\mu} = \frac{e^{-ip_{\mu}a} - e^{ip_{\mu}a}}{2a}
\equiv -i \frac{P_{\mu} + P_{\mu}^{+}}{2} \eqno(2.3)$$
where $p_{\mu}$'s are generators of translation. The forward and the
backward differences may be expressed in terms of the symmetric
difference
$$\delta^{f}_{\mu} = \delta^{s}_{\mu} + \Delta^{a}_{\mu}, \hspace{1in}
\delta^{s}_{\mu} = \delta^{s}_{\mu} - \Delta^{a}_{\mu} \eqno(2.4)$$
where the antisymmetric combination $\Delta_{\mu}^{a}$.
$$\Delta^{a}_{\mu} \equiv \frac{e^{-ip_{\mu}a} + e^{ip_{\mu}a} -
2}{2a} \eqno(2.5)$$
vanishes in the continuum limit.

The `naive' lattice action is the result of sustituting the
symmetric difference $\delta^{s}_{\mu}$ for the derivative
$\partial_{\mu}$ in the continuum euclidean action
$$\begin{array}{lll}
S^{naive} &= &- \displaystyle\sum_{\mu} <
\bar{\psi}\vert\gamma_{\mu}\left(\frac{e^{-ip_{\mu}a} -
e^{ip_{\mu}a}}{2a}\right) \vert \psi > \\
&= &- \displaystyle\sum_{x,\mu} \frac{1}{2a} \left[\bar{\psi}(x + \mu)
\gamma_{\mu}\psi(x) - \bar{\psi}(x)\gamma_{\mu}\psi(x + \mu)\right]
\end{array}\eqno(2.6)$$
where $\mu = ae_{\mu}$ with $e_{\mu}$ a unit vector in the
$\mu$-direction. Within the Brillouin zone $-{\pi/a}  \leq
k_{\mu} \leq {\pi/a}$ the inverse propagator
$$G^{naive}_{F} (k)^{-1} = i\gamma_{\mu} \frac{\sin(k_{\mu}a)}{a}
\eqno(2.7)$$
derived from the naive action vanishes at $k_{\mu} = 0$ and
$k_{\mu} = {\pi/a}$. The entire Brillouin zone, therefore,
splits up into 16 distinct sectors corresponding to the domains
$$0 \leq \vert k_{\mu} \vert \leq \frac{\pi}{2a}, \hspace{1in} \frac{\pi}{2a}
\leq \vert k_{\mu} \vert \leq \frac{\pi}{a} \eqno(2.8)$$
of each component of the four momenta, in each of which the fermion
propagator assumes the standard continuum form
$$G^{naive}_{F}(k)^{-1} = i\gamma_{\mu}^{A} k_{\mu}$$
with $\gamma^{A}_{\mu}(A = 1,2,.... 16)$ unitarily equivalent to
$\gamma_{\mu}$. The physical fermion resides in the first hypercube
$0 \leq \vert k_{\mu} \vert \leq {\pi/2a}$, all $\mu$, and
corresponds to $\gamma^{A}_{\mu} = \gamma_{\mu}$, while the 15
doublers live in the remaining sectors$^{1,4}$.

Appearance of the doublers is not peculiar to the naive action,
but, according to the no-go theorem$^{2}$, is generic. A simple
version of the theorem states$^{3}$ that the 15 doublers are
unavoidable if the lattice inverse propagator has (i) continuity,
(ii) hypercubic symmetry, (iii) reflection symmetry i.e.,
$\gamma_{4}G^{-1}_{F}(k)^{\dagger}\gamma_{4} = G^{-1}_{F}\left(k_{i} -
k_{4}\right)$, and (iv) chiral symmetry $\gamma_{5}
G^{-1}_{F}(k)\gamma_{5} = - G^{-1}_{F}(k)$. Models have been
proposed which eliminate the doublers, either completely or
partially, at the expense of one or more of the above properties.

In Wilson model chiral symmetry is abandoned to achieve complete
elimination of the doublers in the continuum spectrum
$$\begin{array}{lcl}
S^{W} &= &- \displaystyle\sum_{\mu} <
\bar{\psi}\vert\gamma_{\mu}\left(\frac{e^{-ip_{\mu}a} -
e^{ip_{\mu}a}}{2a}\right) + r\left(\frac{e^{-ip_{\mu}a} + e^{ip_{\mu}a} -
2}{2a}\right)\vert\psi > \\
&=
&-\displaystyle\sum_{x,\mu}\frac{1}{2a}\left[\bar{\psi}(x+\mu)(r+\gamma_{\mu})\psi(x)
- \bar{\psi}(x)(r -\gamma_{\mu})\psi(x+\mu)\right. \\
& &\left.- 2r\bar{\psi}(x)\psi(x)\right]
\end{array} \eqno(2.9)$$
The Wilson term makes a mass-like contribution of $0\left({r/a}\right)$
in the inverse propagator
$$G^{W}_{F}(k)^{-1} =
\displaystyle\sum_{\mu}\left[i\gamma_{\mu}\frac{\sin(k_{\mu}a)}{a} + \frac{r}{a} \left(1 -
\cos(k_{\mu}a)\right)\right] \eqno(2.10)$$
in the neighbourhood of $k_{\mu} \approx {\pi/a}$ where the
doublers live. This lifts the degeneracy between the physical
fermion and the doublers, and removes the latter from the spectrum
in the limit $a = 0^{1,4}$.

\section{Chirally Invariant Model for Lattice Fermion}

The species doublers in the naive lattice action (2.6) owe their
origin to the substitution of the symmetric difference for
continuum derivative. A recipe for cure, which follows
naturally$^{7}$, is to use instead forward (2.1) or backward (2.2)
difference. A serious fallout of this recipe is the loss of
hermiticity of the action. The integrand in the partition function
ceases to have the interpretation of a probability measure, and
Feynman rules would yield terms which do not have correct reality
properties in the continuum limit$^4$.

Intimately related to the hermiticity of action is the chiral
structure and anomaly of a gauge theory. Consider the
continuum Dirac operator of a non-abelian chiral gauge theory
$$i D\!\!\!\!/\,_{\pm} (A) \equiv - \left(\partial\!\!\!/\, - ig
\frac{1 \pm \gamma_{5}}{2} A\!\!\!\!/\,\right) \eqno(3.1)$$
with $A_{\mu} = A^{\dagger}_{\mu}$. The effective actions $\Gamma_{+}(A)$
and $\Gamma_{-}(A)$ defined by
$$\Gamma_{\pm}(A) = - \ln\det \left(i D\!\!\!\!/\,_{\pm}(A)\right) \eqno(3.2)$$
are related to each other through hermitian conjugation
$$\Gamma_{+}(A) = \Gamma_{-}(A)^{\dagger} \eqno(3.3)$$
The relation (3.3) is of essence for the chiral properties of the
theory$^{8}$ and implies that chiral anomaly is related to
imaginary part of the effective actions $\Gamma_{\pm}(A)$. Eq.(3.3)
is a direct consequence of the hermiticity of the Dirac operator
$D\!\!\!\!/\,(A)$.
$$\begin{array}{lll}
D\!\!\!\!/\,(A) &\equiv &\left(i \partial\!\!\!/\, + g A\!\!\!\!/\,\right) \\
&= &\left(\begin{array}{cc}
        0 & D(A) \\
        D^{\dagger}(A) &0 \end{array}\right)\end{array} \eqno(3.4)$$
The Weyl components $D(A), D^{\dagger}(A)$
$$D(A) \equiv \sigma_{\mu}\left(i\partial_{\mu} + gA_{\mu}\right) \eqno(3.5)$$
with $\sigma_{\mu} = (i, \sigma_{k})$, are hermitian conjugates of
each other and eq.(3.3) follows from
$$\Gamma_{+}(A) = - \ln \det \left[D(A) D^{\dagger}(0)\right]$$
$$\Gamma_{-}(A) = - \ln \det \left[D(0) D^{\dagger}(A)\right] \eqno(3.6).$$

The naive lattice action (2.6) with gauge field interactions through
link variables  $U_{\mu}$
$$S^{naive} = - \displaystyle\sum_{\mu} < \bar{\psi} \vert \gamma_{\mu}
\left(\frac{e^{-ip_{\mu}a} U^{\dagger}_{\mu} -
U_{\mu}e^{ip_{\mu}a}}{2a}\right) \vert
\psi > \eqno(3.7)$$
obeys the hermiticity condition even for finite lattice spacing. The
fact that it fails to reproduce the ABJ anomaly is well
understood$^{1,4}$ in terms of cancellation of the contribution from
the physical fermion by those from the doublers.

The key elements for a chirally invariant formulation of lattice Dirac
fermion, therefore, are (i) absence of doublers (ii) chiral symmetry,
and (iii) hermiticity. Note that there is no requirement of
hermiticity for the Weyl components of a Dirac operator. All these
elements are realised if in the continuum Dirac operator we substitute
forward (backward) and backward (forward) finite differences for the
derivatives in the right-handed and the left-handed Weyl components
respectively
$$\gamma_{\mu}\partial_{\mu} \rightarrow \left(\begin{array}{cc}
0 &\sigma_{\mu}\left(\frac{e^{-ip_{\mu}a} - 1}{a}\right) \\
\sigma_{\mu}^{\dagger} \left(\frac{1 - e^{ip_{\mu}a}}{a}\right) &0
\end{array}\right) \eqno(3.8)$$
In Dirac basis the wave operator (3.8) assumes the form
$$\gamma_{\mu} \left(\frac{e^{-ip_{\mu}a} - e^{ip_{\mu}a}}{2a}\right) -
\gamma_{\mu}\gamma_{5} \left(\frac{e^{-ip_{\mu}a} + e^{ip_{\mu}a} - 2}{2a}\right)
= \gamma_{\mu}\delta^{s}_{\mu} - \gamma_{\mu}\gamma_{5}\Delta^{a}_{\mu}
\eqno(3.9)$$
The `irrelevant term' $\gamma_{\mu}\gamma_{5}\Delta^{a}_{\mu}$
coincides with the Wilson term in (2.9) if in the latter the parameter
$r$ is replaced by $\gamma_{\mu}\gamma_{5}$. The representation (3.9)
has the correct hermiticity properties, is chirally invariant, and, as
is easily verified, free from doublers.

The price paid$^{3}$ for eliminating the doublers is the loss of
hypercubic and reflection symmetries. Hypercubic symmetry is the
remnant of $SO(4)$ symmetry on lattice and is the invariance under
rotation through ${\pi/2}$. This is lost because the
antisymmetric combination $\Delta^{a}_{\mu}$ does not transform as a
vector. As for reflection symmetry,
$\gamma_{4}\gamma_{5}\Delta^{a}_{4}$ is compatible but the remaining
pieces in the irrelevant term are not. Note that the sum of squares
$\displaystyle\sum_{\mu}(\gamma_{\mu}\gamma_{5}\Delta^{a}_{\mu})^{2}$, does not
suffer from these incompatibilities. This suggests as remedy to use
instead the wave operator
$$\gamma_{\mu}\delta^{s}_{\mu} -
\gamma_{\mu}\epsilon_{\mu}\gamma_{5}\Delta^{a}_{\mu} \equiv
\gamma_{\mu}\delta^{s}_{\mu} -
\gamma^{\epsilon}_{\mu}\gamma_{5}\Delta^{a}_{\mu} \eqno(3.10)$$
where $\epsilon_{\mu} = \pm 1$, with the prescription that all
correlation functions are to be obtained after averaging$^{7}$ over
$\epsilon_{\mu}$. What $\epsilon_{\mu}$ achives is the decoupling of
the finite difference, forward or backward, used for the $\mu$-th
component of the derivative from those used for the other components.
The (anti) correlation of finite differences used for the right-handed
and the left-handed Weyl components is, however, maintained by the
wave operator (3.10), so that the chiral structure (3.4) is preserved. The
wave operator (3.10) leads to the `free' fermion propagator
$$G^{0}_{F}(k)^{-1} = i\gamma_{\mu}\frac{\sin\left(k_{\mu}a\right)}{a} +
\gamma_{\mu}^{\epsilon}\gamma_{5} \frac{1 - \cos \left(k_{\mu}a\right)}{a}
\eqno(3.11)$$

Interactions with gauge fields are introduced, as usual, through the
link variables $U_{\mu}$, and the lattice action corresponding to the
wave operator (3.10) is given by
$$\begin{array}{lll}
S_{F} &=
&-\displaystyle\sum_{\mu}\frac{1}{2a}<\bar{\psi}\vert\left[\gamma_{\mu}
\left(e^{-ip_{\mu}a}U^{\dagger}_{\mu}-U_{\mu}e^{ip_{\mu}a}\right)\right.
\\
& &\left.+\gamma^{\epsilon}_{\mu}\gamma_{5}\left(2-e^{-ip_{\mu}a}U^{\dagger}_{\mu}
-U_{\mu}e^{ip_{\mu}a}\right)-m\right]\vert\psi>
\\
&=
&-\displaystyle\sum_{x,\mu}\frac{1}{2a}\left[\{\bar{\psi}(x+\mu)\gamma_{\mu}U^{\dagger}_{\mu}(x)\psi(x)-\bar{\psi}(x)\gamma_{\mu}U_{\mu}(x)\psi(x+\mu)\}\right.
\\
&
&+\{2\bar{\psi}(x)\gamma^{\epsilon}_{\mu}\gamma_{5}\psi(x)-\bar{\psi}(x+\mu)\gamma^{\epsilon}_{\mu}\gamma_{5}U^{\dagger}_{\mu}(x)\psi(x)
\\
&
&\left.-\bar{\psi}(x)\gamma^{\epsilon}_{\mu}\gamma_{5}U_{\mu}(x)\psi(x+\mu)\}-m\bar{\psi}(x)\psi(x)\right]
\end{array} \eqno(3.12)$$
We have introduced a mass `$m$' for the fermion which vanishes in the
chiral limit.

Ward identities for the U(1) vector and axial vector currents are
derived by requiring invariance of the partition function under the
local transformations
$$\psi(x) \rightarrow e^{i\alpha(x)}\psi(x), \hspace{1in} \bar{\psi}(x) \rightarrow
\bar{\psi}(x) e^{-i\alpha(x)}$$
and
$$\psi(x) \rightarrow e^{i\beta(x)\gamma_{5}}\psi(x), \hspace{1in} \bar{\psi}(x)
\rightarrow \bar{\psi}(x)e^{i\beta(x)\gamma_{5}} \eqno(3.13)$$
respectively. The Jacobians of fermion measure in the lattice fermion
action are trivial$^{9}$ for both the transformations (3.13).
Anomalies, if any, can arise only from lattice artifacts, the
`irrelevant terms'. The Ward identities on lattice, therefore, are
$$\frac{1}{a}<J^{+}_{\mu}(x)-J^{+}_{\mu}(x-\mu)> = \frac{1}{a} <
J^{\epsilon -}_{\mu 5}(x) - J_{\mu 5}^{\epsilon -}(x-\mu)>, \eqno(3.14)$$
$$\frac{1}{a}<J^{+}_{\mu 5}(x) - J^{+}_{\mu 5}(x-\mu)> =
\frac{1}{a}<J^{\epsilon -}_{\mu}(x) - J^{\epsilon -}_{\mu}(x-\mu)>,
\eqno(3.15)$$
where
$$J^{+}_{\mu}(x) \equiv
\frac{1}{2}\left[\bar{\psi}(x)\gamma_{\mu}U_{\mu}(x)\psi(x+\mu)+\bar{\psi}(x+\mu)\gamma_{\mu}U^{\dagger}_{\mu}(x)\psi(x)\right],$$
$$J^{+}_{\mu 5}(x) \equiv
\frac{1}{2}\left[\bar{\psi}(x)\gamma_{\mu}\gamma_{5}U_{\mu}(x)\psi(x+\mu)+\bar{\psi}(x+\mu)\gamma_{\mu}\gamma_{5}U^{\dagger}_{\mu}(x)\psi(x)\right],$$
$$J^{\epsilon -}_{\mu 5}(x) \equiv
\frac{1}{2}\left[\bar{\psi}(x)\gamma^{\epsilon}_{\mu}U_{\mu}(x)\psi(x+\mu)-\bar{\psi}(x+\mu)\gamma^{\epsilon}_{\mu}U^{\dagger}_{\mu}(x)\psi(x)\right],$$
$$J^{\epsilon -}_{\mu 5}(x) \equiv
\frac{1}{2}\left[\bar{\psi}\gamma^{\epsilon}_{\mu}\gamma_{5}U_{\mu}(x)\psi(x+\mu)-\bar{\psi}(x+\mu)\gamma^{\epsilon}_{\mu}\gamma_{5}U^{\dagger}_{\mu}(x)\psi(x)\right].
\eqno(3.16)$$
The right hand sides of eqs.(3.14) and (3.15), if non-zero, are to be
identified as anomalies in U(1) vector and axial vector currents
respectively
$$\begin{array}{lll}
<\partial_{\mu}J_{\mu}(x)> &= &\displaystyle\lim_{a\rightarrow 0}<Y> \\
&= &\displaystyle\lim_{a\rightarrow 0}\frac{1}{a}<J^{\epsilon -}_{\mu 5}(x) -
J^{\epsilon -}_{\mu 5}(x-\mu)> \end{array} \eqno(3.17)$$
$$\begin{array}{lll}
<\partial_{\mu}J_{\mu 5}(x)> &= &\displaystyle\lim_{a\rightarrow 0}<X> \\
&= &\displaystyle\lim_{a\rightarrow 0} \frac{1}{a}<J_{\mu}^{\epsilon -}(x) -
J^{\epsilon -}_{\mu}(x-\mu)> \end{array} \eqno(3.18)$$

\section{U(1) Ward Identities in WCPT}

To evaluate $<X>$ and $<Y>$ it is convenient to start from their
operator representations
$$<Y> = Tr\gamma_{5}<x\vert\left(G_{F}R\!\!\!\!/\,^{\epsilon} +
R\!\!\!\!/\,^{\epsilon}G_{F}\right)\vert x>$$
$$<X> = Tr<x\vert\left(G_{F}R\!\!\!\!/\,^{\epsilon} -
R\!\!\!\!/\,^{\epsilon}G_{F}\right)\vert x> \eqno(4.1)$$
where $R\!\!\!\!/\,^{\epsilon} \equiv
\gamma_{\lambda}\epsilon_{\lambda}R_{\lambda}$ with
$$R_{\lambda} \equiv \frac{2 - U_{\lambda}e^{ip_{\lambda}a} -
e^{-ip_{\lambda}a} U_{\lambda}^{\dagger}}{2a} \eqno(4.2)$$
The fermion propagator is given by
$$G^{-1}_{F} = i\left(D\!\!\!\!/\, - iR\!\!\!\!/\,^{\epsilon}\gamma_{5} - im\right)
\eqno(4.3)$$
where
$$D_{\lambda} = \frac{U_{\lambda}e^{ip_{\lambda}a} -
e^{-ip_{\lambda}a} U^{\dagger}_{\lambda}}{2ia} \eqno(4.4)$$
In (4.1) `Tr' stands for trace over $\gamma$-matrices, and, in
non-abelian gauge theory, also over the symmetry matrices.

In {\it wcpt} the link variable $U_{\lambda} \approx 1 +
iagA_{\lambda}$, so
that for small lattice spacing
$$\begin{array}{lll}
D_{\lambda} &= &\left(s_{\lambda} + gA_{\lambda}\cos(p_{\lambda}a)\right) + 0(a)
\\
R_{\lambda} &= &c_{\lambda} + 0(a) \end{array} \eqno(4.5)$$
where
$$s_{\lambda} =
\frac{\sin\left(p_{\lambda}a\right)}{a}, \hspace{1in} c_{\lambda} = \frac{1 -
\cos(p_{\lambda} a)}{a} \eqno(4.6)$$
The commutators in {\it wcpt} are given by
$$\begin{array}{lll}
\left[R_{\lambda}, R_{\rho}\right] &= &\left[D_{\lambda}, R_{\rho}\right] = 0 \\
\left[D_{\lambda}, D_{\rho}\right] &= &F_{\lambda\rho}\cos(p_{\lambda}a)\cos(p_{\rho}a) 
\equiv f_{\lambda\rho} \end{array} \eqno(4.7)$$
The quadratic form of the propagator ${\cal G}$, defined by
$$\begin{array}{lll}
{\cal G}^{-1} &\equiv &\left(G_{F} G^{\dagger}_{F}\right)^{-1} \\
&= &\left(D\!\!\!\!/\, + iR\!\!\!\!/\,^{\epsilon}\gamma_{5} - im\right)
\left(D\!\!\!\!/\, + iR\!\!\!\!/\,^{\epsilon}\gamma_{5} + im\right) \\
&= &\displaystyle\sum_{\lambda}\left(D_{\lambda}^{2} + R^{2}_{\lambda}\right) +
\frac{i}{2}\sigma_{\lambda\rho}f_{\lambda\rho} -
2\gamma_{5}\sigma_{\lambda\rho}\epsilon_{\rho}R_{\rho}D_{\lambda}+m^{2}
\end{array} \eqno(4.8)$$
commutes with $\gamma_{5}$ and its trace with odd powers of
$\gamma$-matrices vanish. These properties lead to a simpler
structures
for $<Y>$ and $<X>$
$$\begin{array}{lll}
<Y> &= &Tr\gamma_{5}<x\vert\left(D\!\!\!\!/\,{\cal G}R\!\!\!\!/\,^{\epsilon} +
R\!\!\!\!/\,^{\epsilon}{\cal G}D\!\!\!\!/\,\right)\vert x> \\
<X> &= &Tr<x\vert\left(D\!\!\!\!/\,{\cal G}R\!\!\!\!/\,^{\epsilon} -
R\!\!\!\!/\,^{\epsilon}{\cal G}D\!\!\!\!/\,\right)\vert x> - 2i Tr
\gamma_{5}<x\vert R\!\!\!\!/\,^{\epsilon}{\cal G}R\!\!\!\!/\,^{\epsilon}\vert x> \end{array} \eqno(4.9)$$
In {\it wcpt} where both the lattice spacing `$a$' and the gauge
coupling `$g$'
are regarded as small, one can define a `potential' $V$
$${\cal G}^{-1} = {\cal G}^{-1}_{o} + V$$
and develop a perturbative series$^{5}$
$${\cal G} = {\cal G}_{o} - {\cal G}_{o}V{\cal G}_{o} + {\cal
G}_{o}V{\cal G}_{o}V{\cal G}_{o} + ........  \eqno(4.10)$$
where
$${\cal G}^{-1}_{o} = \displaystyle\sum_{\lambda}\left(s^{2}_{\lambda} + c^{2}_{\lambda}\right) +
m^{2} \eqno(4.11)$$
and
$$V =
\displaystyle\sum_{\lambda}\left(2gA_{\lambda}s_{\lambda}\cos\left(p_{\lambda}a\right)+g^{2}A^{2}_{\lambda}\right)
+ \frac{i}{2}\displaystyle\sum_{\lambda, \rho}\sigma_{\lambda\rho}f_{\lambda\rho} -
2\gamma_{5}\displaystyle\sum_{\lambda,
\rho}\sigma_{\lambda\rho}\epsilon_{\rho}R_{\rho}D_{\lambda}\eqno(4.12).$$

Two factors play key role in the calculation of $<X>$ and $<Y>$ : (i)
terms odd in $R_{\lambda}$ drop out because of $\epsilon$-averaging,
and (ii) in the `physical' sector of the loop momentum $0 \leq
\vert k_{\lambda}\vert \leq {\pi/2a}, R_{\lambda}$ is of $0(a)$ whereas in the
`doubler' sector ${\pi/2a} \leq \vert k_{\lambda}\vert \leq {\pi/a}$
it is of $0\left({1/a}\right)$ and behaves like the mass of a Pauli-Villars
regulator field$^{10}$. One obtains (see Appendix)
$$\displaystyle\lim_{a\rightarrow 0} <Y> = 0 \eqno(4.13)$$
so that the U(1) vector current is conserved
$$<\partial_{\mu}J_{\mu}(x)> = 0 \eqno(4.14)$$
In the expression for $<X>$, however, only the first term in (4.9)
vanishes so that
$$\begin{array}{lll}
<X> &= &-2i Tr \gamma_{5} <x\vert R\!\!\!\!/\,^{\epsilon}{\cal G}R\!\!\!\!/\,^{\epsilon}\vert x > \\
&= &2itr\left(F_{\lambda\rho}\tilde{F}_{\lambda\rho}\right)
\int^{\frac{\pi}{a}}_{-\frac{\pi}{a}}\frac{d^{4}k}{(2\pi)^{4}}
\frac{\displaystyle\sum_{\lambda}c^{2}_{\mu}\displaystyle\Pi_{\alpha}
\left(\cos(k_{\alpha}a)\right)}{\left[\displaystyle\sum_{\lambda}(s^{2}_{\lambda}
+c^{2}_{\lambda})+m^{2}\right]^{3}}
\\
&=
&(i/2\pi^{4})tr\left(F_{\lambda\rho}\tilde{F}_{\lambda\rho}\right)
\displaystyle\sum^{4}_{\nu=1}(-1)^{\nu} {^{3}C_{\nu - 1}} I_{\nu} \end{array} \eqno(4.15)$$
where $\tilde{F}_{\lambda\rho}$ is the dual field tensor, `tr' stands
for trace of symmetry matrices, and
$$I_{\nu} = \int^{\frac{\pi}{2a}}_{-\frac{\pi}{2a}} d^{4}k \frac{\left(\frac{1
+ \cos(k_{\mu}a)}{a}\right)^{2}}{\left[\displaystyle\sum^{4}_{\lambda =
1}s^{2}_{\lambda} + \displaystyle\sum^{\nu}_{\lambda = 1}\left(\frac{1 +
\cos(k_{\lambda}a)}{a}\right)^{2}+m^{2}\right]^{3}} \eqno(4.16)$$
In the continuum limit $I_{\nu} = {\pi^{2}/2\nu}$. Thus the
model reproduces the ABJ anomaly
$$\begin{array}{lll}
< \partial_{\mu} J_{\mu 5}(x) > &= &\displaystyle\lim_{a\rightarrow 0} <X> \\
&= &-(i/16\pi^{2}) tr
\left(F_{\lambda\rho}\tilde{F}_{\lambda\rho}\right) \end{array}
\eqno(4.17)$$
Note that in (4.9) the subscript $\nu$ of $I_{\nu}$ has the meaning of
the number of components of the loop momentum with support in the
doubler sector ${\pi/2a} \leq \vert k_{\mu}\vert \leq {\pi/a}$.
For finite `$m$' the contribution from $\nu = 0$ vanishes in the
continuum limit. If, however, $m$ is zero the latter would exactly
cancel the right hand side of (4.15). For a nonvanishing anomaly it is,
therefore, essential that the continuum limit is taken first and
chiral limit $m = 0$, if necessary, afterwards. This is true also for
the calculation of anomaly in the Wilson model$^{5}$.

\section{Concluding Remarks}

Our search for a model of lattice fermion which is, (i) free from
doublers, (ii) hermitian, and (iii) chirally invariant, has paid rich
dividends. Not only is the model free from doublers, it reproduces in
{\it wcpt} in the continuum limit the correct Ward identities, and, most
importantly, the ABJ anomaly in the U(1) axial vector current. In
this respect the model measures up to the Wilson model. In a very
crucial way, however, the two models differ. Whereas chiral symmetry
is broken explicitly by the Wilson term, the irrelevant  term in the
proposed model is chirally invariant. This makes the latter a viable
candidate for chiral lattice gauge theories.

An important lesson of the present exercise is that the ABJ anomaly
can be realised even with a chirally invariant term. What is important
is that the physical fermion has, to start with, a small but non-zero
mass and the chiral limit $m= 0$ is taken only after the continuum
limit. If the order of the two limits is reversed the ABJ anomaly
disappears. The same observation, interestingly enough, is true also
in the chiral symmetry breaking Wilson model. In the absence of the
mass term, a chirally invariant action for lattice gauge theory
breaks up into two completely decoupled Weyl components. In Weyl
basis, vector and axial vector currents lose their individual identity
and both the currents are conserved. By coupling the two Weyl
components, the mass term brings about an asymmetry between the vector
and the axial vector currents. This asymmetry is apparently the
genesis of the ABJ anomaly. The ABJ anomaly, once it materialises,
persists even when the parameter $m$, which triggered it, vanishes.
There is a parallelism here with the phenomenon of ferromagnetism. A
magnetic field, however weak, is needed to trigger the magnetised state of
ferromagnet, and, once realised, the state persists even in the
symmetruy limit corresponding to zero field. For ferromagnetism, we
have to take the thermodynamic limit first and the symmetry limit
afterwards just as for ABJ anomaly the continuum limit $a = 0$ must
precede the chiral limit.

\newpage
\centerline{\bf\large Appendix}

We show in the following how the first term of (4.9), let us call it
$<X_{o}>$, and $<Y>$ vanish in the continuum limit. The crucial
ingredients are Dirac trace, $\epsilon_{\mu}$-averaging and locality of
the matrix elements.

To illustrate the locality property, concentrate on a typical term
$$< x \vert R_{\alpha}{\cal G}_{o}f_{\lambda\rho}{\cal
G}_{o}R_{\lambda}D_{\nu}{\cal G}_{o}D_{\beta}\vert x>$$
$$= \int\int<x\vert c_{\alpha}{\cal G}_{o}\vert
k>\frac{d^{4}k}{(2\pi)^{4}}<k\vert
f_{\lambda\rho}\vert y>d^{4}y<y\vert{\cal G}_{o}R_{\lambda}D_{\nu}{\cal G}_{o}D_{\beta}\vert x>$$
$$= \int\int
\frac{d^{4}k}{(2\pi)^{4}}\left[\frac{e^{ik(x-y)}c_{\alpha}}{\displaystyle\sum_{\lambda}\left(c^{2}_{\lambda}
+ s^{2}_{\lambda}\right) + m^{2}}\right]d^{4}y
f_{\lambda\rho}(y)<y\vert {\cal G}_{o}R_{\lambda}D_{\nu}{\cal G}_{o}D_{\beta}\vert x>$$
In the physical region $0 \leq \vert k_{\mu} \vert \leq
{\pi/2a}, c_{\alpha} \sim 0(a)$ and the integral vanishes in the
continuum limit, whereas in the nonphysical or doubler sector
${\pi/2a} \leq \vert k_{\mu} \vert \leq {\pi/a}$, the
denominator in the square bracket is of $0\left({1/a^{2}}\right)$ but the
numerator is at most of order $0\left({1/a}\right)$ unless $x = y$. The
same argument can be continued and extended for the remaining factors
and hence proves the locality prperty. Note the parallelism here of
the role of $c_{\lambda}$ in doubler sector with that of the mass of
the Pauli-Villars regulator field in the calculation of chiral
anomalies$^{10}$.

We now illustrate (4.13). Using locality and Dirac trace
$$\begin{array}{lll}
<Y> &= &tr\gamma_{5}<x\vert D\!\!\!\!/\,{\cal G}R\!\!\!\!/\,^{\epsilon} +
R\!\!\!\!/\,^{\epsilon}{\cal G}D\!\!\!\!/\,\vert x> \\
&= &-2 tr\gamma_{5}<x\vert R^{\epsilon}_{\alpha}D_{\alpha}{\cal
G}_{o}V{\cal G}_{o}V{\cal G}_{o}\vert x> \end{array}$$
where $V = V_{5}$ or ${i/2}\sigma_{\lambda\rho}f_{\lambda\rho}$.
Averaging of $\epsilon_{\mu}$ and locality now gives
$$\begin{array}{lll}
<Y> &= &-i tr\left(\sigma_{\lambda\rho}\sigma{\lambda\rho}\right)<x\vert
R^{2}_{\alpha}D_{\alpha}{\cal G}_{o}\left(D_{\beta}{\cal
G}_{o}f_{\lambda\rho} + f_{\lambda\rho}{\cal G}_{o}D_{\beta}\right){\cal
G}_{o}\vert x> \\
&= &-i
tr \left(\sigma_{\lambda\rho}\sigma_{\alpha\beta}\right)<x\vert\frac{\left(R^{2}_{\alpha}
+ R^{2}_{\beta}\right)}{2}\left(D_{\alpha}D_{\beta} +
D_{\beta}D_{\alpha}\right)f_{\lambda\rho}{\cal G}^{3}_{o}\vert x> \\
&= &0 \end{array}$$

Next we turn our attention to $<X_{o}>$.
$$<X_{o}> = <\tilde{X}_{o}> - <\tilde{X}_{o}^{\dagger}>,$$
where
$$<\tilde{X}_{o}> = Tr <x\vert D\!\!\!\!/\,{\cal
G}R\!\!\!\!/\,^{\epsilon}\vert x>$$
now,
$$<\tilde{X}_{o}> = <\tilde{X}_{o1}> + <\tilde{X}_{o2}>,$$
with,
$$<\tilde{X}_{o1}> = tr <x\vert R\!\!\!\!/\,^{\epsilon}{\cal
G}_{o}V_{5}{\cal
G}_{o}\frac{i}{2}\sigma_{\lambda\rho}f_{\lambda\rho}{\cal
G}_{o}D\!\!\!\!/\,\vert x>$$
and
$$<\tilde{X}_{o2}> = tr <x\vert R\!\!\!\!/\,^{\epsilon}{\cal
G}_{o}\frac{i}{2}\sigma_{2\rho}f_{2\rho}{\cal G}_{o}V_{5}{\cal
G}_{o}D\!\!\!\!/\,\vert x>,$$
after plugging in the perturbative expansion for ${\cal G}$ in terms of
${\cal G}_{o}$ and $V$.

After $\epsilon_{\mu}$-averaging, putting in ${R^{2}/4}$ in
place of $R^{2}_{\alpha} \left(R^{2} = \displaystyle\sum_{\lambda}R^{2}_{\lambda}\right)$, and
using locality, we finally obtain,
$$<\tilde{X}_{o1}> = \frac{i}{4}\displaystyle\sum_{\beta,\lambda,\nu,\rho}
Tr\left(\gamma_{5}\sigma_{\beta\nu}\sigma_{\lambda\rho}\right)<x\vert
R^{2}D_{\beta}{\cal G}_{o}D_{\nu}{\cal G}_{o}f_{\lambda\rho}{\cal
G}_{o}\vert x>$$
Similarly, we find, $<\tilde{X}_{o2}> = - <\tilde{X}_{o1}>$, so that
$<\tilde{X}_{o}> = 0$, implying $<X_{o}> = 0$ in the continuum limit.

\end{document}